\documentclass[aps,prd,twocolumn,a4paper,superscriptaddress,nofootinbib]{revtex4-1}
\usepackage{amsmath}
\usepackage{amssymb}
\usepackage{graphicx}
\usepackage{multirow}
\usepackage{dcolumn}
\usepackage{bm}
\usepackage{latexsym}
\usepackage[british]{babel}
\usepackage{soul}
\usepackage{nicefrac}
\usepackage[colorlinks,linkcolor=blue,citecolor=blue,urlcolor=blue ]{hyperref}

\begin{document}

\newcommand{\be}{\begin{equation}}
\newcommand{\ee}{\end{equation}}
\newcommand{\bq}{\begin{eqnarray}}
\newcommand{\eq}{\end{eqnarray}}
\newcommand{\bsq}{\begin{subequations}}
\newcommand{\esq}{\end{subequations}}
\newcommand{\bc}{\begin{center}}
\newcommand{\ec}{\end{center}}

\title{Current and Future White Dwarf Mass-radius Constraints on Varying Fundamental Couplings and Unification Scenarios}
\author{D. M. N. Magano}
\email{up201404130@fc.up.pt}
\affiliation{Centro de Astrof\'{\i}sica da Universidade do Porto, Rua das Estrelas, 4150-762 Porto, Portugal}
\affiliation{Faculdade de Ci\^encias, Universidade do Porto, Rua do Campo Alegre, 4150-007 Porto, Portugal}
\author{J. M. A. Vilas Boas}
\email{up201403623@fc.up.pt}
\affiliation{Centro de Astrof\'{\i}sica da Universidade do Porto, Rua das Estrelas, 4150-762 Porto, Portugal}
\affiliation{Faculdade de Ci\^encias, Universidade do Porto, Rua do Campo Alegre, 4150-007 Porto, Portugal}
\author{C. J. A. P. Martins}
\email{Carlos.Martins@astro.up.pt}
\affiliation{Centro de Astrof\'{\i}sica da Universidade do Porto, Rua das Estrelas, 4150-762 Porto, Portugal}
\affiliation{Instituto de Astrof\'{\i}sica e Ci\^encias do Espa\c co, CAUP, Rua das Estrelas, 4150-762 Porto, Portugal}

\date{6 September 2017}

\begin{abstract}
We discuss the feasibility of using astrophysical observations of white dwarfs as probes of fundamental physics. We quantify the effects of varying fundamental couplings on the white dwarf mass-radius relation in a broad class of unification scenarios, both for the simple case of a polytropic stellar structure model and for more general models. Independent measurements of the mass and radius, together with direct spectroscopic measurements of the fine-structure constant in white dwarf atmospheres lead to constraints on combinations of the two phenomenological parameters describing the underlying unification scenario (one of which is related to the strong sector of the theory while the other is related to the electroweak sector). While currently available measurements do not yet provide stringent constraints, we show that forthcoming improvements, expected for example from the Gaia satellite, can break parameter degeneracies and lead to constraints that ideally complement those obtained from local laboratory tests using atomic clocks.
\end{abstract}

\maketitle

\section{\label{intro}Introduction}

Cosmology and particle physics are presently experiencing a truly exciting period. While both have remarkably successful standard models, the observational evidence for the acceleration of the universe shows that these are at the very least incomplete, and that new physics may be there, waiting to be discovered. A key driver for the forthcoming generation of ground and space-based astrophysical facilities is to search for, identify and ultimately characterize this new physics.

Tests of the stability of fundamental couplings---effectively testing the universality of physical laws as we know them---are one of the cornerstones of this endeavor \cite{Uzan,GRG}. High-resolution spectroscopic studies of absorption systems along the line of sight of bright quasars have led to indications of variations of the fine-structure constant at the parts per million level of relative variation \cite{Dipole}, and additional tests of this claim are being actively pursued \cite{LP3,Kotus}. Meanwhile, analogous tests have also been carried out for compact astrophysical objects, including solar-type stars \cite{Adams,stars}, Population III stars \cite{Ekstrom} and neutron stars \cite{Perez}. Typically these yield constraints that are not as strong as the quasar ones (largely due to uncertainties associated with nuclear physics processes), although it is important to note that the two types of tests are carried out in very different physical environments and therefore they are in any case independent tests. In particular, tests in compact astrophysical objects are important to constrain possible dependencies of fundamental couplings on the local environment, e.g. the strength of the local gravitational field.

In this work we will discuss how another class of astrophysical objects, white dwarfs, can be used for similar purposes. The physics of white dwarfs is comparatively well known \cite{Shapiro}, and it can be constrained through their mass-radius relation which is a convenient observable. While the number of objects for which the mass and radius have been independently determined with good accuracy and without assuming any underlying model is currently small \cite{holberg}---of order ten, to be compared to the tens of thousands of known white dwarfs---this number and the sensitivity of the measurements are both expected to increase in the coming years, for example as a result of the Gaia space mission.

Indeed the white dwarf mass-radius relation has been recently used to constrain a class of modified gravity models \cite{jain}. Here we show that the standard mass-radius relation is affected in models with spacetime variations of fundamental couplings. In this respect white dwarfs are particularly promising, because spectroscopic measurements of the value of the fine-structure constant $\alpha$ and the proton-to-electron mass ratio $\mu$ can be made on their surface \cite{barrow,Bagdonaite}. For this reason they provide a further tool with which one may be able to constrain models with environmental dependencies \cite{Olive,Marvin}; such constraints have already been obtained locally, relying on the varying gravitational potential felt by the Earth as it orbits the Sun \cite{Clock1,Clock2}. In this work we will consider two such models, without and with explicit environmental dependencies.

We will work in the context of a broad class of Grand Unified Theory (GUT) models, where the variations of the relevant couplings are related in a particular way \cite{coc,Campbell}. This class of models has also been considered in previous works on solar type and neutron stars \cite{stars,Perez}, and it can also be constrained in laboratory tests of the stability of fundamental couplings using atomic clocks \cite{relogios}. Indeed, we will show how constraints on these models coming from white dwarfs can ideally complement the atomic clock ones.

\section{\label{unif}Phenomenology of Unification}

In order to account for the effects of varying fundamental couplings on white dwarfs we must describe phenomenologically the simultaneous variation of the relevant couplings. These couplings will clearly include the fine-structure constant $\alpha=e^2/(\hbar c)$, the proton-to-electron mass ratio $\mu=m_p/m_e$ and Newton's gravitational constant $G$. The simplest way to do this is to relate the various changes to those of a particular dimensionless coupling, typically $\alpha$. Then, if $\alpha=\alpha_0(1+\delta_\alpha)$ and
\begin{equation}
\frac{\Delta A}{A}=k_A\, \frac{\Delta\alpha}{\alpha} \,,
\end{equation}
we have $A=A_0(1+k_A\delta_A)$ and so forth; these $k_A$ are known as sensitivity coefficients. Clearly the relations between the various couplings will be model-dependent. Here we will adopt the generic class of unification models developed in Coc {\it et al.} \cite{coc}, to which we refer the reader for full derivations. Earlier less generic scenarios have also been discussed in \cite{Langacker,Calmet,Dent}.

Specifically, \cite{coc} consider a class of GUT models in which the weak scale is determined by dimensional transmutation and further assuming that relative variations of all the Yukawa couplings $h_i$ are the same, in other words that
\be
\frac{\Delta h_i}{h_i}=\frac{\Delta h}{h}\,.
\ee
Finally they also assume that the variation of the couplings is driven by a dilaton-type scalar field, as described in \cite{Campbell}. In this case one finds that the variations of $\alpha$ and those of other quantities are related through two dimensionless paramters, $R$ and $S$, defined as
\be
\frac{\Delta\Lambda}{\Lambda}=R\frac{\Delta\alpha}{\alpha}+({\rm Electroweak\, terms})\,,
\ee
where $\Lambda$ denotes the energy scale of Quantum Chromodynamics, and
\be
\frac{\Delta v}{v}=S\frac{\Delta h}{h}\,,
\ee
where $v$ is the Higgs vacuum expectation value and $h$ are the aforementioned Yukawa couplings (assumed to have a common relative variation). In this case one can show \cite{coc} that the proton-to-electron mass ratio $\mu$ obeys
\begin{equation}\label{defmualpha}
\frac{\Delta\mu}{\mu}=[0.8~R-0.3~(1+S)]\frac{\Delta\alpha}{\alpha}\,.
\end{equation}

Note that different models within this class will have different values of $R$ and $S$. Their absolute value can be anything from order unity to several hundreds, but while $R$ can be positive or negative (with the former case being more likely), physically one expects that $S\ge0$. To give just two examples, Coc {\it et al.} \cite{coc} suggest typical values of $R\sim36$ and $S\sim160$, while in the dilaton-type model studied by Nakashima {\it et al.} \cite{Nakashima} we have $R\sim109$ and $S\sim0$. We mention these both to illustrate that they are calculable in particular models within this class, and that there is a significant model dependence in their values. Additional discussion of these points can be found in the review by Uzan \cite{Uzan}. In any case, we can simply treat both as phenomenological parameters to be constrained by astrophysical data. The strongest current constraints on $R$ and $S$---or, strictly speaking, a combination thereof---come from atomic clock tests \cite{relogios}.

Concerning the gravitational constant $G$, we must bear in mind that speaking of variations of dimensional constants has no physical significance {\it per se}: one can always concoct any variation one wishes by defining appropriate units of length, time and energy. Still, one is free to choose an arbitrary dimensionful unit as a standard and compare it with other quantities. If one explicitly or implicitly assumes particle masses to be constant, then constraints on $G$ are in fact constraining the (dimensionless) product of $G$ and the nucleon mass squared. A better route is to compare the strong interaction with the gravitational one: this can be done by assuming a fixed energy scale for Quantum Chromodynamics (QCD) and allowing a varying $G$, or vice-versa.

Here, we will follow the latter route, defining the dimensionless couplings
\begin{equation}\label{def_alphas}
\alpha_i=\frac{Gm^2_i}{\hbar c}
\end{equation}
and assuming that the QCD scale and particle masses vary, while the Planck mass is fixed. We then have for the electron mass
\begin{equation}
\frac{\Delta\alpha_e}{\alpha_e}=2\frac{\Delta m_e}{m_e}
   =(1+S)~\frac{\Delta\alpha}{\alpha}
\end{equation}
while for the proton mass
\begin{equation}\label{alphap}
\frac{\Delta\alpha_p}{\alpha_p}=2\frac{\Delta m_p}{m_p}=2 \big[0.8~R+0.2 (1+S) \big]~\frac{\Delta\alpha}{\alpha}\,.
\end{equation}
The combination of the last two equations trivially recovers Eq. \ref{defmualpha}. Similarly for the mass difference between neutrons and protons, $\sigma=m_n-m_p$, we find
\begin{equation}
\frac{\Delta \sigma}{\sigma}=[0.1+0.7~S-0.6~R]\, \frac{\Delta\alpha}{\alpha}\,,
\end{equation}
while for the ratio $\eta=m_n/m_p$ we have
\begin{equation}
\frac{\Delta \eta}{\eta}=\left(\frac{1}{\eta}-1\right)
   \left[0.1-0.5~S+1.4~R\right]\,\frac{\Delta\alpha}{\alpha}\,,
\end{equation}
so the variation is of higher order (i.e., smaller), since the function of $R$ and $S$ is in this case multiplied by a prefactor of order $10^{-3}$. Similarly the relative variation of the neutron mass can be obtained from
\begin{equation}
\frac{\Delta m_n}{m_n}=\frac{\Delta\sigma}{\sigma}+\frac{m_p}{m_n}\left(\frac{\Delta m_p}{m_p}-\frac{\Delta\sigma}{\sigma} \right)  \,.
\end{equation}
Relative variations of other quantities of interest, such as the neutron lifetime and the deuteron binding energy can also be cast in this form, as discussed in \cite{coc}.

Finally, for our present purposes it is also useful to define an average nucleon mass
\be
m_N=\frac{1}{2}(m_p+m_n)\,,
\ee
as well as its corresponding dimensionless parameter $\alpha_N$. However, it is straightforward to show that the difference between the relative variation of $\alpha_N$ and $\alpha_p$ is also of higher order
\be
\frac{\Delta\alpha_N}{\alpha_N}-\frac{\Delta\alpha_p}{\alpha_p}=\frac{2\eta}{1+\eta}\frac{\Delta\eta}{\eta}=\frac{\eta-1}{\eta+1}\left(\frac{\Delta\sigma}{\sigma}-\frac{\Delta m_p}{m_p}\right)\,,
\ee
and therefore we will later make the approximation $\alpha_N\sim\alpha_p$.

\section{\label{polytropic}Mass-radius relation for polytropic white dwarfs}

A polytropic star is a simplified model for the structure of a star in equilibrium \cite{Chandra}. We will start by briefly reviewing the physics underlying the model and using it for discussing the case of a polytropic white dwarf allowing for varying couplings in this section because, having an analytic solution, it will help to understand the more general model (which can only be solved numerically) which we will subsequently study. In this and the following section we follow the canonical textbook treatments of Chandrasekhar \cite{Chandra} and Shapiro and Teukolsky \cite{Shapiro}, to which we refer the reader for further details.

\subsection{Polytropic stars}

The model for a polytropic star starts from the mass continuity equation
\begin{equation}
\dfrac{dm(r)}{dr}= 4 \pi \rho(r) r^2 \:,
\label{mass_continuity}
\end{equation}
and further assumes perfect spherical symmetry and hydrostatic equilibrium
\begin{equation}
\dfrac{dP(r)}{dr}=  - \dfrac{Gm(r) \rho(r)}{r} \:.
\label{hydro_equilibrium}
\end{equation}
Here $r$ is the radial distance to the center of the star, $m$ is the mass within the sphere of radius $r$, $\rho$ the density, and $P$ the pressure. These equations can be solved if an equation of the type $P=P(\rho)$ is specified. The polytropic solution corresponds to a simplified equation of state with the form
\begin{equation}
P= K \rho^{1+\nicefrac{1}{n}} \:,
\label{poly_eq_state}
\end{equation}
where $K$ is the polytropic constant (related to the boundary conditions of the star) and $n$ is called the polytropic index and is in principle a free parameter. From these equations we can arrive at the Lane-Emden equation, given by
\begin{equation}
\dfrac{1}{z^2} \dfrac{d}{dz} \Big( z^2 \dfrac{dw}{dz} \Big) = - w^n \:.
\label{laneeden}
\end{equation}
In order to obtain this equation, we have made the following substitutions ($\rho_c$ is the density at the center of the star):
\begin{equation}
z=\dfrac{r}{a} \:,
\qquad
w = \Big( \dfrac{\rho}{\rho_c} \Big)^{\nicefrac{1}{n}} \:,
\qquad
a^2 =\dfrac{(1+n)K}{4 \pi G {\rho_c}^{1-\nicefrac{1}{n}}} \: .
\label{substitutions}
\end{equation}
The corresponding solution will have physical significance from $z=0$ until $z=z_n$, with $z_n$ being the first zero of the parameter $w$ for the chosen value of $n$.

If we define $R_{\star}$ and $M_{\star}$ to be the total radius and mass of the star, respectively, then
\begin{equation}
\dfrac{(1+n)K}{(4 \pi)^{\nicefrac{1}{n}}} \beta_n = R_{\star}^{\nicefrac{3}{n}-1} M_{\star}^{1-\nicefrac{1}{n}} \:,
\label{poly_relation}
\end{equation}
where $\beta_n$ is a factor that can be calculated numerically:
\begin{equation}
\beta_n= z_n^{1+\nicefrac{1}{n}} (-w'(z_n))^{1-\nicefrac{1}{n}} \:.
\label{beta}
\end{equation}

\subsection{Simplified mass-radius relation}

A white dwarf is a low- or medium-mass star in the final stage of its life, after the main sequence. Having burned up all the nuclear fuel, the thermal pressure can no longer support its own gravity. Hydrostatic equilibrium is achieved because electrons become degenerate, and the resulting Fermi pressure prevents the star from collapsing. Here we start by considering a simple model of a white dwarf using the free electron gas model, and assuming Newtonian gravity and no thermal effects, following the analysis of \cite{Shapiro}.

For free electrons the number of states $dn$ available at momentum $p$ per unit volume is
\begin{equation}
dn =\dfrac{p^2 dp}{\pi^2 \hslash^3} \: .
\label{dn}
\end{equation}
The electrons will occupy one octant of a sphere of radius $p_{F}$ in the $p$-space, whose volume is
\begin{equation}
\dfrac{1}{8} \dfrac{4 \pi}{3} p_{F} ^3=\dfrac{1}{2}qN \dfrac{\hslash^3 \pi^3}{V} \: ,
\label{pf}
\end{equation} 
where $p_{F}$ is the Fermi momentum, $N$ is the number of nucleons in the gas, $q$ is the number of electrons per nucleon and $V$ is the total volume of the gas. 

Firstly, we explore the situation where the electrons are non-relativistic. In this case, the energy of the system is given by 
\begin{equation}
E_{NR}=\int_{0}^{p_{F}} \dfrac{p^2}{2 m_{e}} \dfrac{V p^2 dp}{\pi^2 \hslash^3} = \dfrac{\hslash^2 (3 \pi^2 N q)^{\nicefrac{5}{3}}}{10 \pi^2 m_{e} V^{\nicefrac{2}{3}}} \: ,
\label{Enr}
\end{equation} 
where $m_{e}$ is the mass of the electron. From here can we find the value of the degeneracy pressure:
\begin{equation}
P_{NR} = -\Big( \dfrac{\partial E}{\partial V} \Big)_{N} =  \dfrac{(3\pi^2)^{\nicefrac{2}{3}}}{5} \dfrac{\hslash^2}{m_e} \Big(\dfrac{q \rho}{m_N} \Big)^{\nicefrac{5}{3}}\: ,
\label{Pnr}
\end{equation} 
where $m_N$ is the mass of the nucleon and $\rho$ is the density of the material. Indeed, we are assuming that all the mass comes from the atomic nuclei:
\begin{equation}
\rho = \dfrac{N m_N}{V}\: .
\label{rho}
\end{equation} 

We can also consider the ultra-relativistic limit $E \approx pc$:
\begin{equation}
E_{UR}=\int_{0}^{p_{F}} pc \dfrac{V p^2 dp}{\pi^2 \hslash^3} = \dfrac{(3^2 \pi)^{\nicefrac{2}{3}}}{4} \dfrac{\hslash c (Nq)^{\nicefrac{4}{3}}}{V^{\nicefrac{1}{3}}} \:,
\label{Eur}
\end{equation} 
\begin{equation}
P_{UR} = \dfrac{(3 \pi^2)^{\nicefrac{1}{3}}}{4} \hslash c \Big( \dfrac{q \rho}{m_N} \Big)^{\nicefrac{4}{3}}\:.
\label{Pur}
\end{equation} 

Equations \eqref{Pnr} and \eqref{Pur} provide us with polytropic equations of state for the non-relativistic and ultra-relativistic white dwarf, respectively. Combining the former with equation \eqref{poly_relation}, we find a theoretical mass-radius relation for the non-relativistic white dwarf:
\begin{equation}
R_{\star} = \Big( \dfrac{3 \pi}{64} \Big)^{\nicefrac{2}{3}} \beta_{\nicefrac{3}{2}} (2q)^{\nicefrac{5}{3}} \Big( \dfrac{G \hslash^2}{c^4} \Big)^{\nicefrac{2}{3}} \dfrac{1}{{\alpha_N}^{\nicefrac{5}{6}} {\alpha_e}^{\nicefrac{1}{2}}} \dfrac{1}{M_{\star}^{\nicefrac{1}{3}}} \:.
\label{mass-radius_NR}
\end{equation} 
It is also useful to rewrite this equation in terms of dimensionless masses and radii. We then have
\begin{equation}
{\cal R}_{\star}= \Big( \dfrac{3 \pi}{64} \Big)^{\nicefrac{2}{3}} \beta_{\nicefrac{3}{2}} (2q)^{\nicefrac{5}{3}} \dfrac{1}{{\alpha_N}^{\nicefrac{5}{6}} {\alpha_e}^{\nicefrac{1}{2}}} \dfrac{1}{{\cal M}_{\star}^{\nicefrac{1}{3}}} \:,
\label{mass-radius_NRdimensionless}
\end{equation}
where ${\cal R}_{\star}$ and ${\cal M}_{\star}$ are in units of Planck's radius and mass, respectively.

If we now consider the ultra-relativistic regime, we get the Chandrasekhar limit (again, in units of the Planck mass)
\begin{equation}
{\cal M}_{\star}=\sqrt{\dfrac{3 \pi}{64}} {\beta_3}^{\nicefrac{3}{2}} (2q)^2 \dfrac{1}{\alpha_N}\:.
\label{mass-radius_UR}
\end{equation} 
Here the $\alpha_i$ are defined in equation \eqref{def_alphas}, with $m_i$ referring to the masses of the nucleon and the electron.

Now we are in a position to apply the modifications mentioned in Sect. \ref{unif} to our model (which would otherwise be standard). We replace
\begin{equation}
\dfrac{1}{{\alpha_N}^{\nicefrac{5}{6}} {\alpha_e}^{\nicefrac{1}{2}}} \rightarrow \dfrac{1-x}{{\alpha_p}^{\nicefrac{5}{6}} {\alpha_e}^{\nicefrac{1}{2}}} \:,
\qquad
\dfrac{1}{\alpha_N} \rightarrow \dfrac{1-y}{\alpha_p}\:,
\label{substituicao_poli}
\end{equation}
where we have used the fact that the relative variations of $\alpha_N$ and $\alpha_p$ differ by higher-order terms (cf. Sect. \ref{unif}) to replace the former by the latter, and for convenience we have defined
\begin{eqnarray}
x= \Big[ \dfrac{4}{3}R+\dfrac{5}{6}(1+S) \Big]  \dfrac{\Delta \alpha}{\alpha}  \:,
\label{x_poli} \\
y= \Big[ \dfrac{8}{5}R+\dfrac{2}{5}(1+S) \Big]  \dfrac{\Delta \alpha}{\alpha} \:.
\label{y_poli}
\end{eqnarray}
In short, our equations for the white dwarf have the following structure
\begin{eqnarray}
R_{\star}=\dfrac{0.0126}{M_{\star}^{\nicefrac{1}{3}}}\,(1-x) \qquad \text{(non-relativistic)} \:
\label{mass_radius_NR_final} \\
M_{\star}=1.45\,(1-y) \qquad \text{(ultra-relativistic)} \:,
\label{mass_radius_UR_final}
\end{eqnarray} 
where the numerical values apply for $R_{\star}$ and $M_{\star}$ expressed in units of solar radius and mass, respectively.

\section{\label{non_poly}General Mass-Radius Relation}

We will now discuss a more general model, not restricting ourselves to any of the relativistic limits. As a consequence, we will no longer be able to find a simple analytic expression for the mass-radius relation. As we shall see, the behavior of this model will differ significantly from the previous one, especially for white dwarfs with large masses.

First, it is convenient to introduce the dimensionless quantity
\begin{equation}
x_F=\dfrac{p_F}{m_e c}= {\Big(\dfrac{3 \pi^2 (2q)}{2 m_N} \Big)}^{\nicefrac{1}{3}} \dfrac{\hslash^2}{m_e c} \rho^{\nicefrac{1}{3}} \:.
\label{x_f}
\end{equation}
The energy of the star is
\begin{eqnarray}
\begin{aligned}
E {} & = \int_{0}^{p_{F}} (\sqrt{{m_e}^2 c^4+p^2 c^2}-m_e c^2) \dfrac{V p^2 dp}{\pi^2 \hslash^3} \\
& = \dfrac{V {(m_e c^2)}^4}{\hslash^3 c^3 \pi^2} \int_{0}^{x_F} (\sqrt{1+x^2}-1)x^2 dx \\
& =  \dfrac{V {(m_e c^2)}^4}{\hslash^3 c^3 \pi^2} \zeta(x_F) \:.
\label{energia_WD}
\end{aligned}
\end{eqnarray} 
In order to find the pressure, it is easier to calculate the integral over the flux of momentum:
\begin{eqnarray}
\begin{aligned}
P {} & = \int_{0}^{p_{F}} \dfrac{vp}{3} \dfrac{p^2 dp}{\pi^2 \hslash^3} \\
&= \dfrac{{m_e}^4 c^5}{3 \pi^2 \hslash^3}  \int_{0}^{x_F} \dfrac{x^4}{\sqrt{1+x^2}} dx \\
&= \dfrac{{m_e}^4 c^5}{3 \pi^2 \hslash^3} \xi(x_F) \:.
\label{pressao_WD}
\end{aligned}
\end{eqnarray} 
The functions $\zeta$ and $\xi$ are
\begin{eqnarray}
\zeta(x)=\dfrac{1}{8\pi^2} \Big[\big(x+2x^3\big)\sqrt{1+x^2}-\log\big(\sqrt{1+x^2}+x\big)\Big] \:, \qquad
\label{zeta} \\
\xi(x)=\dfrac{1}{8\pi^2} \Big[\big(2x^3/3-x\big)\sqrt{1+x^2}+\log\big(\sqrt{1+x^2}+x\big)\Big] \:. \qquad
\label{xi}
\end{eqnarray}

Solving the system of equations \eqref{mass_continuity}, \eqref{hydro_equilibrium} , \eqref{x_f} and \eqref{pressao_WD} provides the functions $m(r)$, $P(r)$, $\rho(r)$ and ${x_F}(r)$, from which we can obtain the mass-radius relation. The surface of the white dwarf is the value of $r$ for which $P(r)=0$ (as previously mentioned we call it $R_{\star}$). It should also be noted that $x=0$ is the only root of $\xi(x)$---c.f. Eq. \eqref{xi}. Therefore, by Eq. \eqref{pressao_WD}, the task of finding $R_{\star}$ is equivalent to the one of finding the (first) root of $x_F(r)$. Naturally, the mass of star is $m(R_{\star}) \equiv M_{\star}$. With this in mind, we conveniently reduce that system to the following one
\begin{equation}
\begin{cases}
\dfrac{dm'}{dr}=m_0 r^2 {x_F}^3\\
\dfrac{dx_F}{dr}=-K_1 \dfrac{m'}{r^2} \dfrac{\sqrt{1+x_F^2}}{x_F}\\
m=K_2 m'
\end{cases} .
\label{eq_system}
\end{equation}
There is no analytic solution to this system, and so we will have to resort to numerical methods. For this purpose, we have introduced the dimensionless constant $m_0$ in order to control the order of magnitude of the parameters in the equations. We also defined
\begin{equation}
K_1= \dfrac{16}{3 \pi (2q)^2} \dfrac{R_{\odot}}{m_0} \dfrac{c^3}{G \hslash^3} \alpha_e \alpha_N \:, \\
\label{k1}
\end{equation}
\begin{equation}
K_2= \dfrac{8}{3 \pi (2q)8} \dfrac{{R_{\odot}}^3}{M_{\odot}m_0} \dfrac{c^5}{G^2 \hslash} {\alpha_e}^{\nicefrac{3}{2}} {\alpha_N}^{\nicefrac{1}{2}} \:.
\label{k2}
\end{equation}

To allow for the possibility of varying couplings, both $K_1$ and $K_2$ should now be extended to include corrections. Making again use of phenomenological relations of Sect. \ref{unif} we get
\begin{equation}
\begin{cases}
\dfrac{dm'}{dr}=m_0 r^2 {x_F}^3\\
\dfrac{dx_F}{dr}=-K_1 (1+\beta) \dfrac{m'}{r^2} \dfrac{\sqrt{1+x_F^2}}{x_F}\\
m=K_2 (1+\gamma) m'
\end{cases} ,
\label{eq_system_final}
\end{equation}
where
\begin{eqnarray}
\beta= \bigg[ \dfrac{9}{5}R+\dfrac{8}{5}(1+S) \bigg] \dfrac{\Delta \alpha}{\alpha} \:,
\label{beta_npoli} \\
\gamma= \bigg[ \dfrac{4}{5}R+\dfrac{23}{10}(1+S) \bigg] \dfrac{\Delta \alpha}{\alpha} \:.
\label{gama_npoli}
\end{eqnarray}
Note that $\beta$ and $\gamma$ are significantly different from the analogous parameters in the polytropic case, $x$ and $y$. 

\begin{figure}
\begin{center}
\includegraphics[width=3.5in]{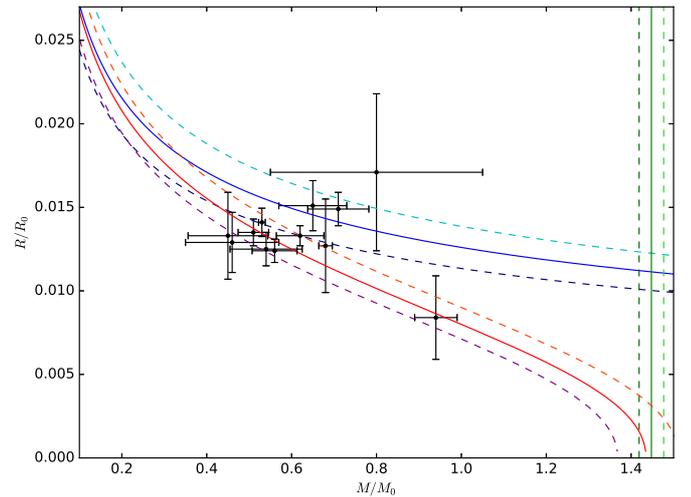}
\caption{\label{r_m_hob}The white dwarf mass-radius relation for the general model (Eq. \protect\eqref{eq_system_final}, solid red line for the standard model, with the nearby darker and lighter dashed lines corresponding to $\beta=\pm0.01$). For comparison the plot also shows the non-relativistic limit of the polytropic model (Eq. \protect\eqref{mass_radius_NR_final}, blue solid line for the standard case and darker and lighter dashed lines for $x=\pm0.1$) and the ultra-relativistic model (Eq. \protect\eqref{mass_radius_UR_final}, green solid line for the standard case and darker and lighter dashed lines for $y=\pm0.02$). The black points with error bars correspond to the data in Table \protect\ref{HOB}.} 
\end{center}
\end{figure}

Indeed, it is instructive to compare this model with the polytropic ones. In Fig. \ref{r_m_hob}, we plot the curves corresponding to Eqs. \eqref{mass_radius_NR_final}, \eqref{mass_radius_UR_final} and \eqref{eq_system_final}. This plot suffices to show that the polytropic models do not tell the whole story about white dwarfs. We can see that Eq. \eqref{mass_radius_NR_final} is an accurate model only for low mass stars. On the other hand, Eq. \eqref{mass_radius_UR_final} is only good for stars very near the Chandrasekhar limit. For these reasons, in the following sections we will base our analysis on Eq. \eqref{eq_system_final}.

\begin{table}
\caption{\label{HOB}Catalog of currently white dwarf masses and radii, reproduced from \protect\cite{holberg}. All masses and radii are in units of $M_{\odot}$ and $R_{\odot}$, respectively, and their values and corresponding uncertainties are shown with the same number of significant digits as the original reference.}
\begin{center}
\begin{tabular}{ |c|c|c| }
 \hline
System & $M_{\star}/M_{\odot}$ & $R_{\star}/R_{\odot}$ \\
 \hline
WD0413$-$077 & $0.51 \pm 0.036$  & $0.0135 \pm 0.0008$ \\
WD0416$-$594 & $0.62 \pm 0.056$  & $0.0133 \pm 0.0006$ \\
WD0642$-$166 & $0.94 \pm 0.05 $  & $0.0084 \pm 0.0025$ \\
WD1105$-$048 & $0.45 \pm 0.094$  & $0.0133 \pm 0.0026$ \\
WD1143$+$321 & $0.71 \pm 0.072$  & $0.0149 \pm 0.001 $ \\
WD1314$+$293 & $0.80 \pm 0.25 $  & $0.0171 \pm 0.0047$ \\
WD1327$-$083 & $0.53 \pm 0.0079$ & $0.0141 \pm 0.00085$ \\
WD1620$-$391 & $0.68 \pm 0.016$  & $0.0127 \pm 0.0028$ \\
WD1706$+$332 & $0.54 \pm 0.085$  & $0.0125 \pm 0.001 $ \\
WD1716$+$020 & $0.65 \pm 0.08 $  & $0.0151 \pm 0.0015$ \\
WD1743$-$132 & $0.46 \pm 0.11 $  & $0.0129 \pm 0.0018$ \\
WD2341$+$322 & $0.56 \pm 0.053$  & $0.0124 \pm 0.0007$ \\
 \hline
\end{tabular}
\end{center}
\end{table}

Note that in this section we are assuming a model for the relative variation of $\alpha$ (and therefore for the other quantities related to it) where its numerical value is the same for all white dwarfs, and is spatially homogeneous on the scale of the white dwarf radius. An alternative scenario, where the magnitude of the (relative) variations does depend explicitly on the local gravitational field, is briefly discussed in the Appendix. We also note that to other astrophysical processes, such as rotation or magnetic fields, may affect the mass-radius relation: as in the recent \cite{jain} (which also uses this to constrain modified gravity models) such additional effects have been neglected, but they may be relevant for future 
datasets.

\section{\label{obs_constrains}Current observational Constraints}

We now use our mass-radius relation model to set constraints on the parameters $\beta$ and $\gamma$. We will make use of a catalog of twelve white dwarfs in binary systems, compiled in \cite{holberg}, for which both masses and radii have been independently obtained from a combination of observations of trigonometric parallaxes, spectroscopic effective temperatures and surface gravities, and gravitational redshifts. These are listed in Table \ref{HOB} and are also depicted by the black points in Fig. \ref{r_m_hob}. We note that a possible source of model dependence in the analysis of \cite{holberg} stems from the fact that their analysis requires an estimate of the the intrinsic flux of the white dwarf, which is made by fitting the observed spectrum to model atmosphere codes. Quantitatively estimating the magnitude of the effect of $\alpha$ variations on this spectroscopic fitting is beyond the scope of this work (as it would require detailed simulations of these spectra) though we believe that this effect is negligible in our current error budget.

We carry out a standard likelihood analysis, with $\beta$ and $\gamma$ as fitting parameters, which is otherwise similar to the one in \cite{jain}, which recently used the same data to constrain a class of modified gravity models. For each star $i$ in our catalog, we choose the value of $M_{\star}$ that minimizes the following quantity
\begin{equation}
{\chi_i}^2 (M_{\star}) = \dfrac{{(M_{\star}-M_i)}^2}{\sigma_{M,i}^2}+\dfrac{{(R_{th}(M_{\star})-R_i)}^2}{\sigma_{R,i}^2} \:,
\label{chi_i}
\end{equation}
where $M_i$, $\sigma_{M,i}$, $R_i$, and $\sigma_{R,i}$ are the mass and radius of the $i$th star and their respective uncertainties, and $R_{th}(M)$ is the theoretical relation that we wish to fit. Thus, total value of $\chi^2$ is 
\begin{equation}
\chi^2 = \sum_{i=1}^{N} {\chi_i}^2(\hat{M_i}) \:,
\end{equation}
with $\hat {M_i}$ as the value of $M$ that minimizes the corresponding ${\chi_i}^2$.

\begin{figure}
\begin{center}
\includegraphics[width=3.5in]{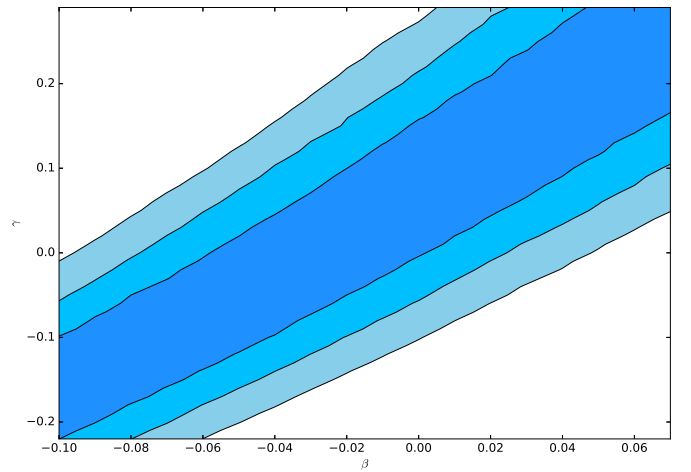}
\caption{\label{no_kepler}One, two and three sigma confidence regions in the $\beta$--$\gamma$ plane, from the currently available data listed in Table \protect\ref{HOB}.}
\end{center}
\end{figure}

Figure \ref{no_kepler} shows the resulting constraints in the $\beta$--$\gamma$ plane. The results are consistent with the standard values $\beta=\gamma=0$, but the two parameters are strongly correlated, preventing us from obtaining individual constraints on each of them. This is partially because the uncertainties in the masses and radii are relatively large, but also due to the fact that the available measurements span a comparatively narrow range of white dwarf masses, around 0.6 solar masses. (The only white dwarf in the catalog with a mass near one solar mass is WD0642$-$166, otherwise known as Sirius B.) Nevertheless, this suggests that the degeneracy between the two parameters can be broken by improved (future) astrophysical measurements, as we will show in what follows.

\section{\label{forecast}Forecasting future constraints}

Astrophysical facilities such as the Gaia satellite should soon lead to significantly improved measurements of white dwarf masses and radii. Here, by means of a simple forecast, we show that these improvements can be expected to lead to competitive constraints on $\beta$ and $\gamma$, as well as on the previously introduced class of unification models.

\begin{figure}
\begin{center}
\includegraphics[width=3.5in]{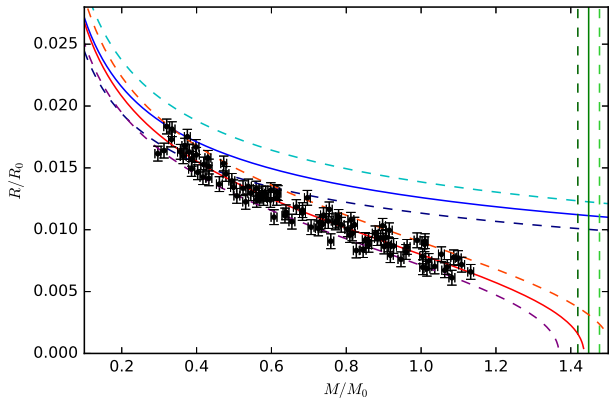}
\caption{\label{r_m_kepler}Same as Fig. \protect\ref{r_m_hob}, except that the black data points now depict a simulated future dataset, as described in the text.}
\end{center}
\end{figure}

For this purpose we have generated a simulated catalog of 100 mass-radius pairs, spanning a wider range of masses, $0.3<M_{\star}<1.2$. We conservatively assume that the fiducial model is the standard one (with $\beta=\gamma=0$). For the scatter of the measurements and their uncertainties, we make the simplifying assumption that each of the corresponding masses and radii is determined with an uncertainty corresponding to the smallest of the currently available ones, which are listed in Table \ref{HOB}. This corresponds to $\sigma_M=0.0079$ and $\sigma_R=0.0006$ respectively, and is likely to be a conservative assumption, both in terms of uncertainties and (perhaps even more so) in terms of the number of measurements---as was pointed out in the introduction, several tens of thousands of white dwarfs are already known. This simulated dataset is plotted against the theoretical mass-radius relations in Fig. \ref{r_m_kepler}, which should be compared to Fig. \ref{r_m_hob}.

\subsection{Constraints on $\beta$ and $\gamma$}

\begin{figure}
\begin{center}
\includegraphics[width=3.5in]{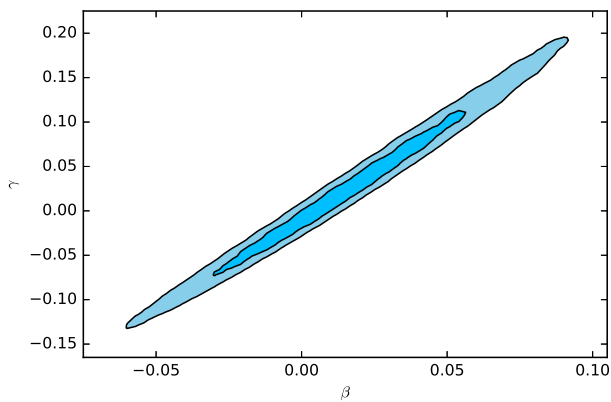}
\caption{\label{kepler}Two and three sigma confidence regions in the $\beta$--$\gamma$ plane, for the simulated future dataset described in the text. To be compared to the current constraints, depicted in Fig. \protect\ref{no_kepler}.}
\end{center}
\end{figure}
\begin{figure}
\begin{center}
\includegraphics[width=3.5in]{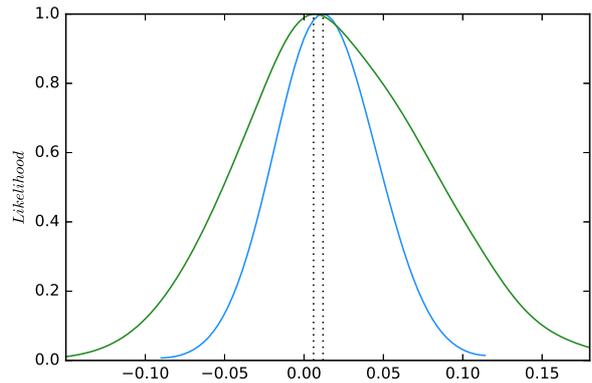}
\caption{\label{likelihood_beta_gama}One-dimensional marginalized likelihoods for $\beta$ and $\gamma$ parameters (blue and green curves respectively) for the simulated future dataset. The vertical dotted lines identify the best-fit values of each parameter.}
\end{center}
\end{figure}

Figure \ref{kepler} shows the constraints obtained from the simulated data on the $\beta$--$\gamma$ plane. The previous degeneracy is partially broken, so marginalized constraints on $\beta$ and $\gamma$ can now be obtained. Figure \ref{likelihood_beta_gama} shows the likelihoods for each parameter, and at the $68.3\%$ (1$\sigma$) confidence levels we find, for the parameters defined in Eqs. \eqref{beta_npoli}-- \eqref{gama_npoli} the following constraints
\begin{eqnarray}
\label{beta_constrains}
\beta=0.012\pm0.032 \:, \\
\label{gamma_constrains}
\gamma=0.006\pm0.060\:.
\end{eqnarray}
Thus each of the parameters can be constrained to an accuracy of a few percent. As a simple illustration, if we assume the typical values suggested in \cite{coc} of $R\sim30$ and $S\sim160$, allowing a 10$\%$ uncertainty in each of them, we find
\begin{equation}
\dfrac{\Delta \alpha}{\alpha}=(2.7\pm9.1) \times 10^{-5} \,,
\end{equation}
which although weaker that direct spectroscopic constraints is consistent with them (and also consistent with the null result, as it should be given our choice of fiducial model).

\subsection{Constraints on $R$ and $S$}

Our parameters $\beta$ and $\gamma$ are specific combinations of the unification parameters $R$ and $S$ and the relative variation of $\alpha$ itself. Therefore, measurements of $\Delta\alpha/\alpha$ on the surface of white dwarfs can be used as priors in this analysis, allowing us to express the direct constraints on $\beta$ and $\gamma$ as constraints on $R$ and $S$ (or possibly combinations thereof). Recently Berengut {\it et al.} \cite{barrow} reported on spectroscopic Hubble Space Telescope measurements in the white dwarf G191-B2B, using Fe V and Ni V transitions which are comparatively very sensitive to $\alpha$ variations, and finding respectively
\begin{equation}
\left(\dfrac{\Delta \alpha}{\alpha}\right)_{\rm Fe V}=(4.2\pm1.6) \times 10^{-5}
\end{equation}
\begin{equation}
\left(\dfrac{\Delta \alpha}{\alpha}\right)_{\rm Ni V}=(-6.1\pm5.8) \times 10^{-5}\,;
\end{equation}
note that the two measurements are discrepant at 1.6 standard deviations; given the high resolution of the spectra used in the analysis, the most likely source of uncertainty is the accuracy of the laboratory wavelength measurements of the required Fe V and Ni V transitions. In what follows we will therefore use them separately in the following analysis, but for comparison we will also consider their weighted mean combination, which is
\begin{equation}
\left(\dfrac{\Delta \alpha}{\alpha}\right)_{\rm Joint}=(3.5\pm1.5) \times 10^{-5}\,.
\end{equation}

\begin{figure*}
\begin{center}
\includegraphics[width=3.5in]{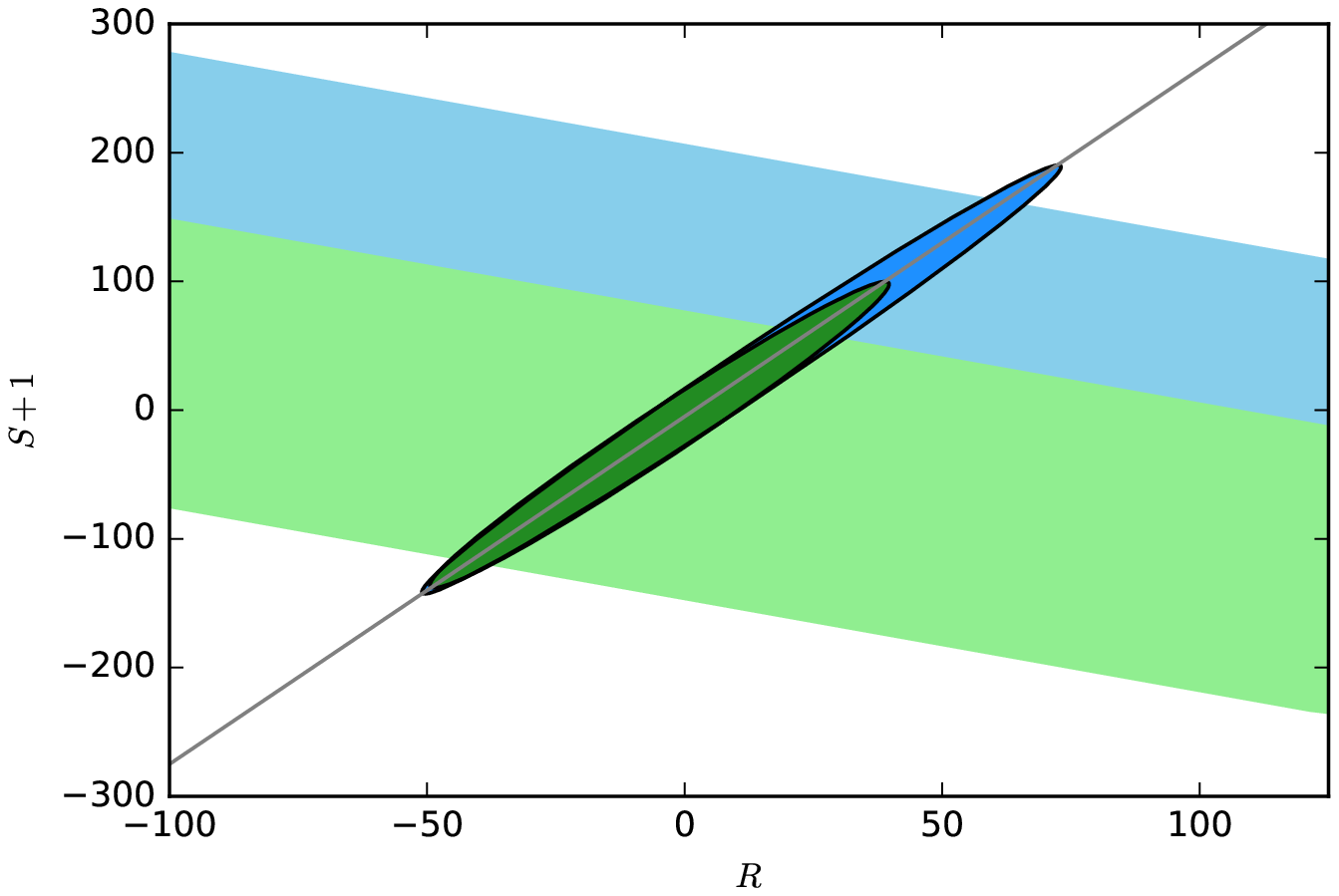}
\includegraphics[width=3.5in]{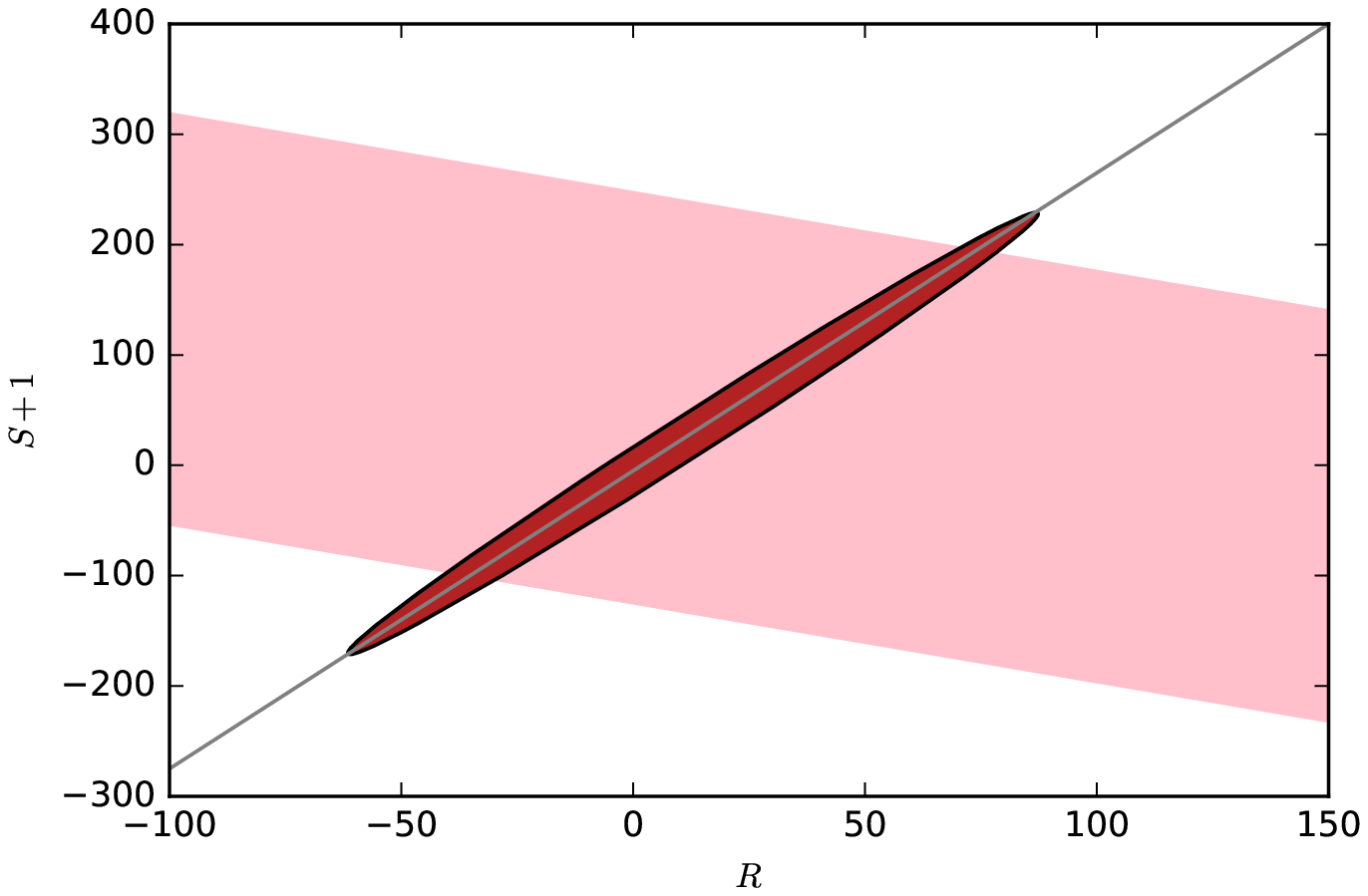}
\caption{\label{R_S}{\bf Left panel:} Allowed one-sigma regions in the $R$--$S$ parameter space. The blue (top) and green (bottom) bands correspond to the Iron and Nickel measurements, respectively. The thin grey band corresponds to the atomic clocks bound, Eq. \eqref{constrain_atomclocks}, while the darker elliptic regions are the result of the combination of the latter with each of the two former ones. {\bf Right panel:} Analogous plot, with the pink band corresponding to the weighted mean combination of the Iron and Nickel measurements.}
\end{center}
\end{figure*}

\begin{figure*}
\begin{center}
\includegraphics[width=3.5in]{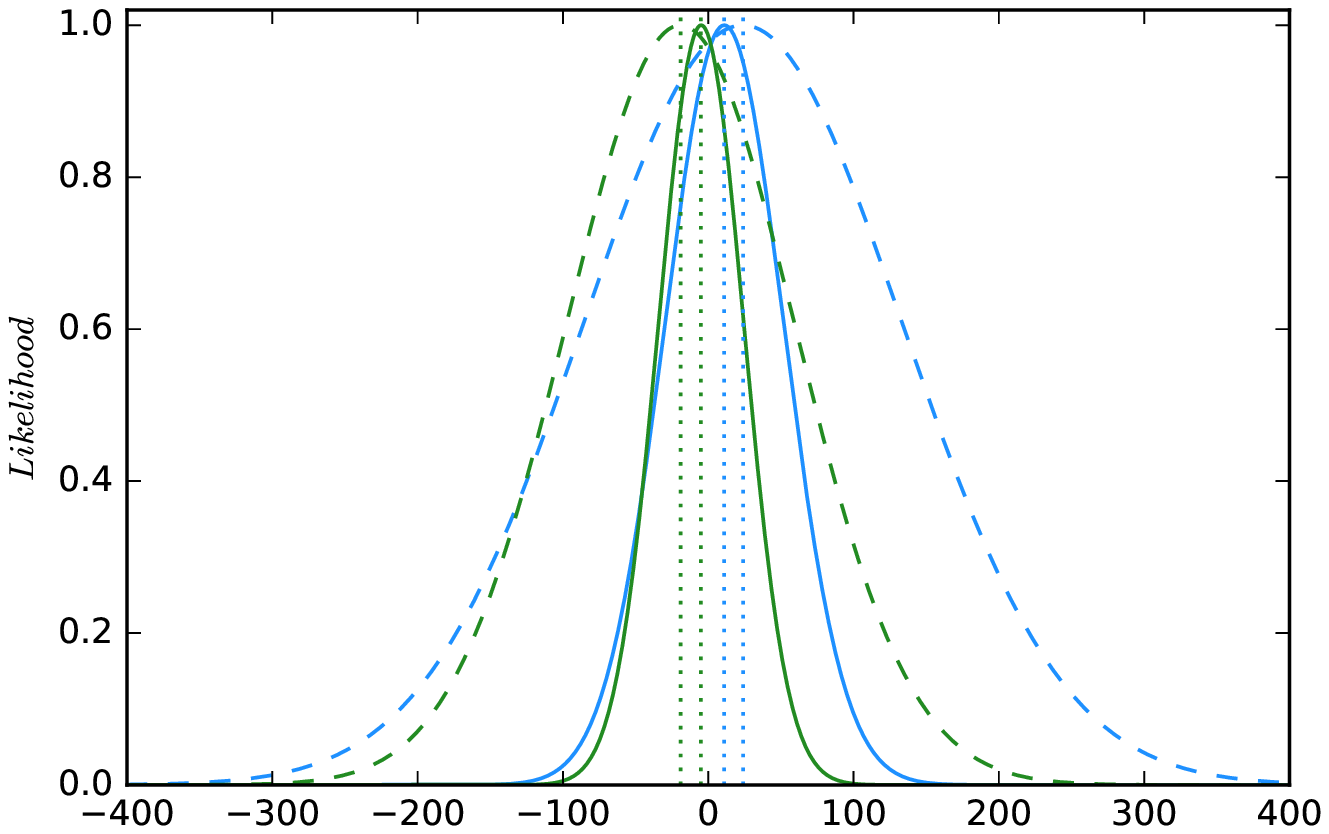}
\includegraphics[width=3.5in]{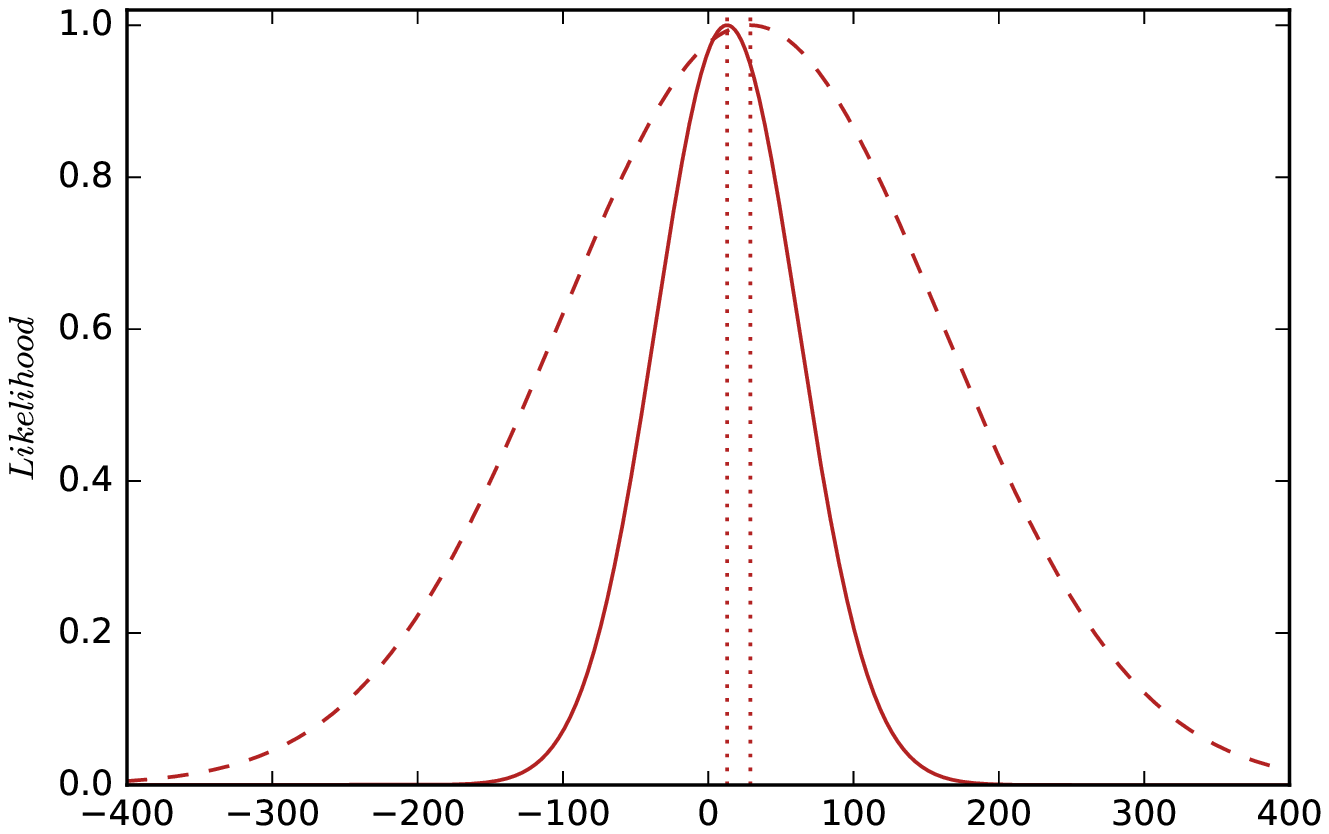}
\caption{\label{1Dlikelihood}One-dimensional posterior likelihoods for $R$ (solid lines) and $S$ (dashed lines), marginalizing over the other parameter. In the left panel the blue lines correspond to the Iron measurements while the green ones correspond to the Nickel measurements; in the right panel the reld lines correspond to the weighted mean of both measurements.}
\end{center}
\end{figure*}

With these measurements we could use Eqs. \eqref{beta_constrains}--\eqref{gamma_constrains} to constrain the two combinations of $R$ and $S$ given by Eqs. \eqref{beta_npoli}--\eqref{gama_npoli}. Note than in this case one can only constrain combinations of the two parameters, rather than each one of them individually. Interestingly, the combination of the parameters $R$ and $S$ to which the white dwarf mass-radius relation is sensitive is orthogonal to an analogous constraint obtained from laboratory tests of the stability of fundamental couplings using atomic clocks \cite{relogios}
\begin{equation}
\label{constrain_atomclocks}
(S+1)-2.7R=-5 \pm 15 \,.
\end{equation}
We can therefore take our forecast one step further and combine the white dwarf and atomic clock constraints, again using a standard likelihood analysis. Figure \ref{R_S} shows the resulting constraints in the $R$--$S$ plane, while Fig. \ref{1Dlikelihood} shows the overall 1D posterior likelihoods for each of the parameters, marginalizing the other. We show the results of the analysis for both the Iron and Nickel measurements taken separately, and for the combination of the two.

\begin{table}
\caption{\label{finalRS}One-sigma posterior uncertainties on the parameters $R$ and $S$, marginalizing over the other parameter, form the combination of white dwarf and atomic clock data. We separately show the results of the analysis for the cases where the Iron and Nickel measurements, or the weighted mean of both measurements, are used as priors.}
\begin{center}
\begin{tabular}{ |c|c|c| }
 \hline
Prior & $R$ ($68.3\%$ C.L.) & $S$ ($68.3\%$ C.L.) \\
 \hline
Fe V measurement only & $11 \pm 41$  & $24 \pm 109$ \\
Ni V measurement only & $-5 \pm 30$  & $-19 \pm 79$ \\
\hline
Joint (weighed mean) & $13 \pm 49$  & $29 \pm 132$ \\
 \hline
\end{tabular}
\end{center}
\end{table}

The resulting constraints on $R$ and $S$ are summarized in Table \ref{finalRS}. We note the different signs of the best-fit values of $R$ and $S$ that this simulated analysis leads to: positive for Iron (and the joint analysis) and negative for Nickel. From a purely theoretical point of view, the former set of parameters would be somewhat natural than the latter one, but in any case all results are consistent with one another within one standard deviation. This ultimately stems from the fact that these constraints are dominated by the atomic clocks data, cf. Eq. (\ref{constrain_atomclocks}). Therefore, despite the simplifying assumptions made in this forecast, it does show that with the expected improvements in the sensitivity of both the white dwarf mass-radius relation and the spectroscopic measurements of $\alpha$ at their surface (for which \cite{barrow} suggests that improvements by up to two orders of magnitude are within reach) this has the potential to become a powerful probe of unification scenarios.

\section{\label{conclusion}Conclusion}

There is growing interest in using compact astrophysical objects as a probe of fundamental physics paradigms. In this work we have focused on white dwarfs, for which there are three relevant observables. Their masses and radii can be measured independently (i.e., without critically relying on theoretical models) in binary systems, while the value of the fine-structure constant $\alpha$ in their atmosphere can be measured spectroscopically.

By studying how the mass-radius relation is affected in a broad class of GUT models where both $\alpha$ and the particle masses are allowed to vary, we have shown that the combination of these observables can lead to constraints on the phenomenological parameters characterizing the unification models. Interestingly, in the space of these phenomenological parameters, constraints coming from white dwarfs are roughly orthogonal to those coming from atomic clock tests \cite{relogios}. 

After showing that the effects of varying couplings are different in a simple polytropic model and in a more detailed model, we have used current as well as simulated data (representative of future observations) to obtain constraints on the relevant parameters. Currently available data consists of only twelve mass-radius pairs \cite{holberg}, with relatively large uncertainties and in a relatively narrow range of masses, and this implies that at the moment no stringent constraints can be obtained: only a degenerate combination of the relevant parameters is constrained. As for published spectroscopic measurements of $\alpha$, they have been done in a single white dwarf \cite{barrow}, though with discrepant results for the two species used, Iron and Nickel.

The number and the sensitivity of the mass-radius measurements are both expected to increase significantly in the near future. In particular, it is expected that the Gaia space mission \cite{Gaia1,Gaia2} will provide highly accurate independent measurements of masses and radii for several hundreds of white dwarfs---our forecast conservatively assumed 100. On the other hand, the current limiting factor in the sensitivity of the spectroscopic measurements is the uncertainty in laboratory measurements of the relevant atomic transitions; should these be improved, the sensitivity can be improved by up to two orders of magnitude. We thus expect that this method will soon provide competitive constraints.

\begin{acknowledgments}

We are grateful to Jiting Hu, John Webb and Kepler de Souza Oliveira for helpful discussions on the subject of this work.

This work was done in the context of project PTDC/FIS/111725/2009 (FCT, Portugal), with additional support from grant UID/FIS/04434/2013. CJM is supported by an FCT Research Professorship, contract reference IF/00064/2012, funded by FCT/MCTES (Portugal) and POPH/FSE (EC).

\end{acknowledgments}

\section*{Appendix: Environmental dependencies}

In the main text we have assumed that the value of $\alpha$ is the same for all the white dwarfs, while this value need not be the standard one (known from lower density environments such as the Earth). In what follows we will very briefly discuss a different scenario. We will assume that the relative variation of $\alpha$ depends on the local gravitational field,
\begin{equation}
\frac{\Delta \alpha}{\alpha} \propto g\,.
\end{equation}
Clearly, in this model there is an explicit environmental dependence: the value of $\alpha$ will have a radial dependence for each white dwarf, and will also have a different value at the surface of each one of them. For a spherically symmetric mass distribution and Newtonian gravity, we will have
\begin{equation}
\label{environalpha}
\frac{\Delta \alpha}{\alpha} =  a_0 \frac{m(r)}{r^2}\,,
\end{equation}
where $a_0$ is a dimensionless constant and $r$ and $m(r)$ are in solar units as in the main text. The system of equations analogous to that of Eq. (\ref{eq_system_final}) is now
\begin{eqnarray}
\begin{cases}
\dfrac{dm'}{dr}=m_0 r^2 {x_F}^3\\
\dfrac{dx_F}{dr}=-K_1 \big(1+\beta_0 \dfrac{m}{r^2} \big) \dfrac{m'}{r^2} \dfrac{\sqrt{1+x_F^2}}{x_F}\\
m=K_2 \big(1+\gamma_0 \dfrac{m}{r^2} \big) m'
\end{cases}
\end{eqnarray}
and our model parameters (analogous to $\beta$ and $\gamma$) are
\begin{equation}
\beta_0= \bigg[ \dfrac{9}{5}R+\dfrac{8}{5}(1+S) \bigg] a_0
\end{equation}
\begin{equation}
\gamma_0= \bigg[ \dfrac{4}{5}R+\dfrac{23}{10}(1+S) \bigg] a_0\,.
\end{equation}
Comparing these with Eqs. \eqref{beta_npoli}--\eqref{gama_npoli} we note that the dependencies on $R$ and $S$ are exactly the same, but instead of being multiplied by the (previously assumed constant) relative variation of $\alpha$ they are now multiplied by the dimensionless constant $a_0$, defined in Eq. \eqref{environalpha}.

\begin{figure}
\begin{center}
\includegraphics[width=3.5in]{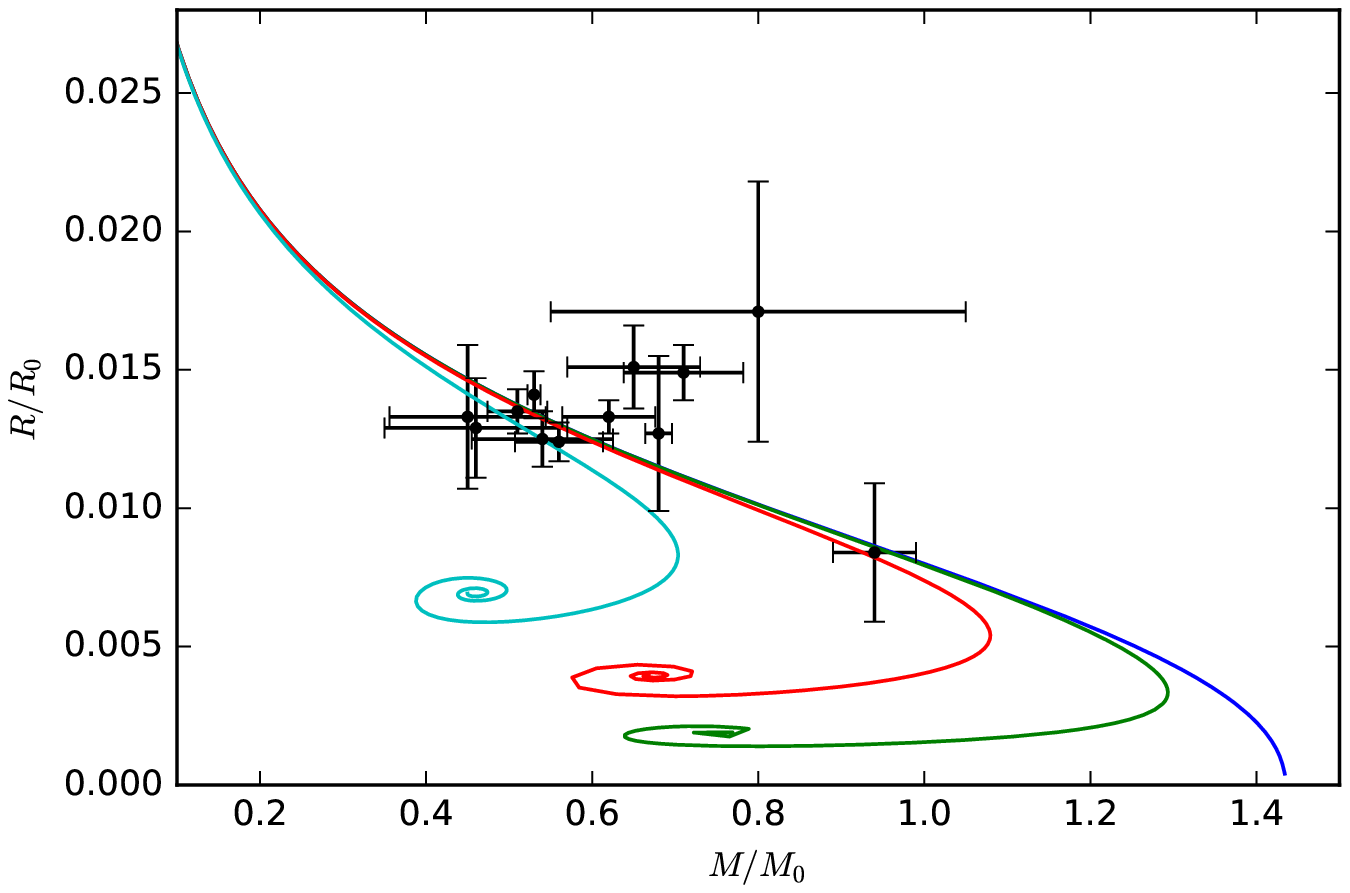}
\includegraphics[width=3.5in]{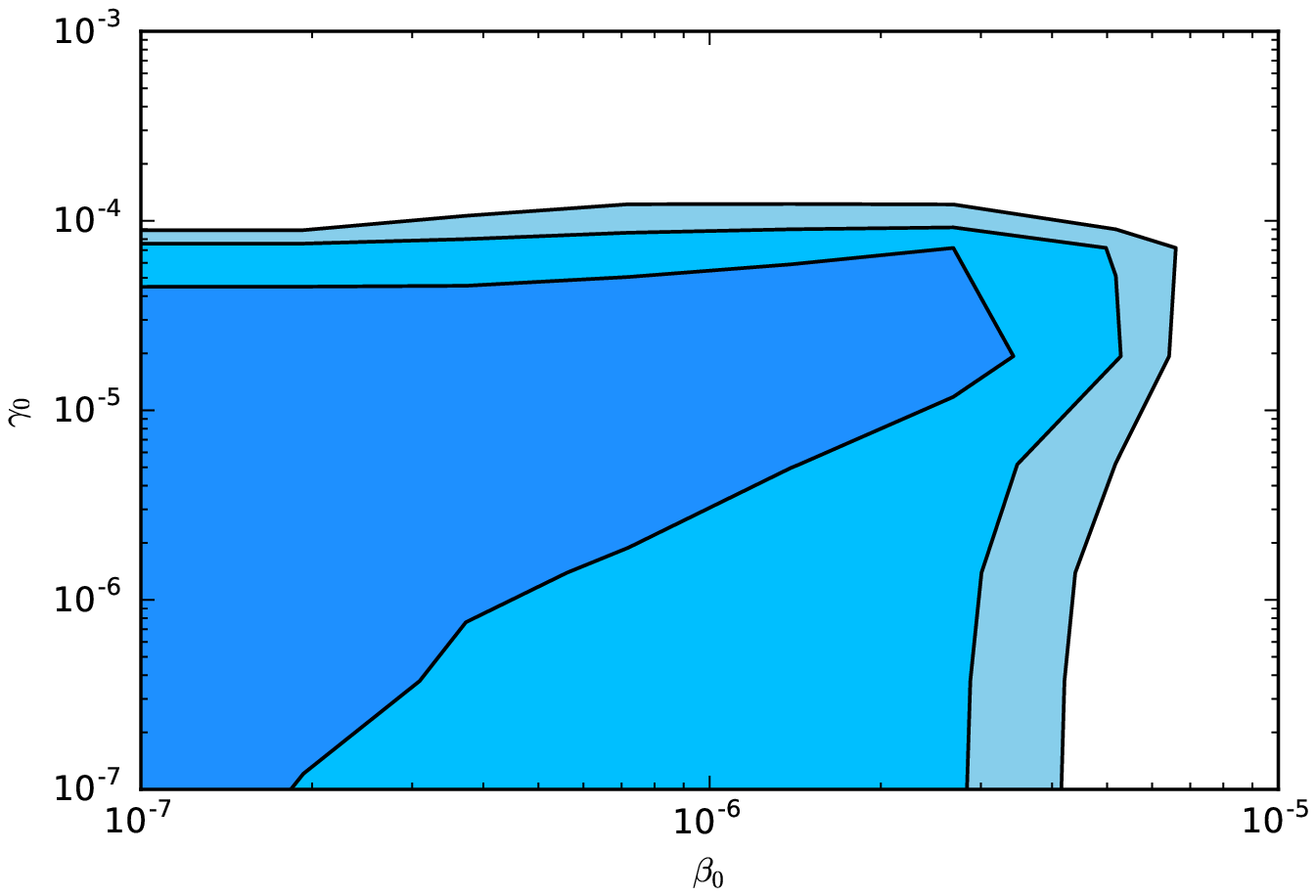}
\caption{\label{figapp}{\bf Top panel:} The white dwarf mass-radius relation for the model where $\alpha$ depends on the local gravitational potential. From left to right, the colored curves correspond to choices of $\beta_0$ and $\gamma_0$equal to $10^{-5}$ $10^{-6}$ $10^{-7}$ and $0$. The black points with error bars correspond to the data in Table \protect\ref{HOB}. {\bf Bottom panel}: One, two and three sigma confidence regions in the $\beta_0$--$\gamma_0$ plane, from the currently available data listed in Table \protect\ref{HOB}.}
\end{center}
\end{figure}

Now there is no advantage in working with $m'$, since it is not obtained from $m$ by a single rescaling. We therefore have
\begin{eqnarray}
\begin{cases}
\dfrac{dm}{dr}=m^2 \Big( \big( \dfrac{1}{m}+\dfrac{\gamma_0}{r^2} \big)^2 K_2 m_0 r^2 {x_F}^3 -\dfrac{2 \gamma_0}{r^3}  \Big) \\
\dfrac{dx_F}{dr}=-\dfrac{K_1}{K_2} \dfrac{1+\beta_0 \dfrac{m}{r^2}}{1+\gamma_0 \dfrac{m}{r^2}} \dfrac{m}{r^2} \dfrac{\sqrt{1+x_F^2}}{x_F}\,.
\end{cases}
\end{eqnarray}
Figure \ref{figapp} shows examples of mass-radius relations for various choices of the parameters $\beta_0$ and $\gamma_0$, and the constraints from the Holberg {\it et al.} \cite{holberg} data listed in Table \ref{HOB}, assuming logarithmic priors for both of these. In this case we find three sigma upper bounds on $\beta_0$ and $\gamma_0$, respectively at the $10^{-5}$ and $10^{-4}$ level. On the other hand, in this case there is no independent determination of $a_0$ (which in principle is another free parameter), so one can't obtain separate constraints on $R$ and $S$.

\bibliography{whitedwarfs}
\end{document}